\documentclass[a4paper]{jpconf}
\usepackage{graphicx}
\usepackage[numbers,square,sort&compress]{natbib}
\begin{document}

\title{Magnetic and physical properties of new hexagonal compounds PrPt$_4X$ ($X$ = Ag, Au)}

\author{Michael O Ogunbunmi and Andr\'{e} M Strydom}

\address{Highly Correlated Matter Research Group, Physics Department, University of Johannesburg, P. O. Box 524,  Auckland Park 2006, South Africa.}

\ead{mogunbunmi@uj.ac.za, moogunbunmi@gmail.com}

\begin{abstract}
	We have synthesized PrPt$_4$Ag and PrPt$_4$Au compounds for the first time and report their crystal structure, as well as magnetic and physical properties in the temperature range of 1.9~K to 300~K. Both compounds are derived from the substitution of Pt with Ag and Au, respectively in the parent compound PrPt$_5$ which crystallizes in the hexagonal CaCu$_5$-type structure. Here we observed the preservation of the hexagonal CaCu$_5$-type structure under such substitutions, which is in contrast to the observations in PrCu$_4$Ag and PrCu$_4$Au adopting the cubic MgCu$_4$Sn-type structure upon substitution on parent hexagonal PrCu$_5$. The temperature dependences of specific heat $C_p(T)$, and electrical resistivity $\rho(T)$ of PrPt$_4$Ag show an anomaly at 7.6~K. This is absent in the magnetic susceptibility $\chi(T)$, thus suggesting a possible multipolar ordering of the Pr$^{3+}$ moment. PrPt$_4$Au on the other hand does not show any anomaly but an upturn in $C_p(T)/T$ below about 10~K and attains 1.23~J/(mol K$^2$) at 1.9~K. In addition, $\rho(T)$ $\sim T$ for  nearly a decade in temperature. These observations for PrPt$_4$Au are the hallmark of a non-Fermi liquid (nFL) behavior and is characteristic of a system near a quantum critical point. The analyses of the low-temperature $C_p(T)$ for PrPt$_4$Ag and PrPt$_4$Au give values of the Sommerfeld coefficient, $\gamma = 728.5~$mJ/(mol K$^2$) and 509.1~mJ/(mol K$^2$), respectively, indicating a significant enhancement of the quasiparticle mass in the two compounds.
\end{abstract}

\section{Introduction}
PrPt$_5$ is a Van-Vleck paramagnet that has been studied mostly for nuclear cooling by adiabatic demagnetization \cite{andres1970nuclear}. This technique profits from the high nuclear spin ($I$ = 5/2) of the only known stable $^{141}$Pr isotope. It crystallizes in the hexagonal CaCu$_5$-type structure with space group $P$6/$mmm$ (No. 191)  \cite{buschow1972note}. It has been observed that the lack of magnetic ordering in PrPt$_5$ is associated with its weak exchange interaction \cite{narasimhan1973low}. In addition, magnetization studies have revealed a nuclear spontaneous ferromagnetic Weiss temperature, $\theta_p$ = 2~mK, which is attributed to nuclear hyperfine enhancement in the system \cite{karaki2000hyperfine}. In a substitution study by Malik {\em et al.} \cite{malik1990structural}, the $R$Pt$_4$In ($R$ = La-Tm) compounds, which represent a doping of 20\% on the Pt site, were reported to adopt the cubic MgCu$_4$Sn-type structure with the exception of TmPt$_4$In which forms in the Cu$_3$Au-type structure. Also, ErPt$_4$In and HoPt$_4$In were observed to possess a mixture of MgCu$_4$Sn and Cu$_3$Au phases as evidenced from the X-ray diffraction patterns. PrPt$_4$In was reported to remain paramagnetic down to 4.2~K \cite{malik1990structural}. Here we report the synthesis, as well as the magnetic and physical properties of the new compounds PrPt$_4$Ag and PrPt$_4$Au. Interestingly, it was found that both compounds crystallize in the same hexagonal CaCu$_5$ parent structure. This is in contrast with the observations in PrCu$_4$Ag and PrCu$_4$Au \cite{zhang2009heavy,zhang2010magnetic} which adopt the cubic MgCu$_4$Sn-type structure similar to those of $R$Pt$_4$In compounds reported by Malik {\em et al.} \cite{malik1990structural}.

\section{Experimental methods}
\begin{figure}[!t]
	\centering
	\includegraphics[scale=0.45]{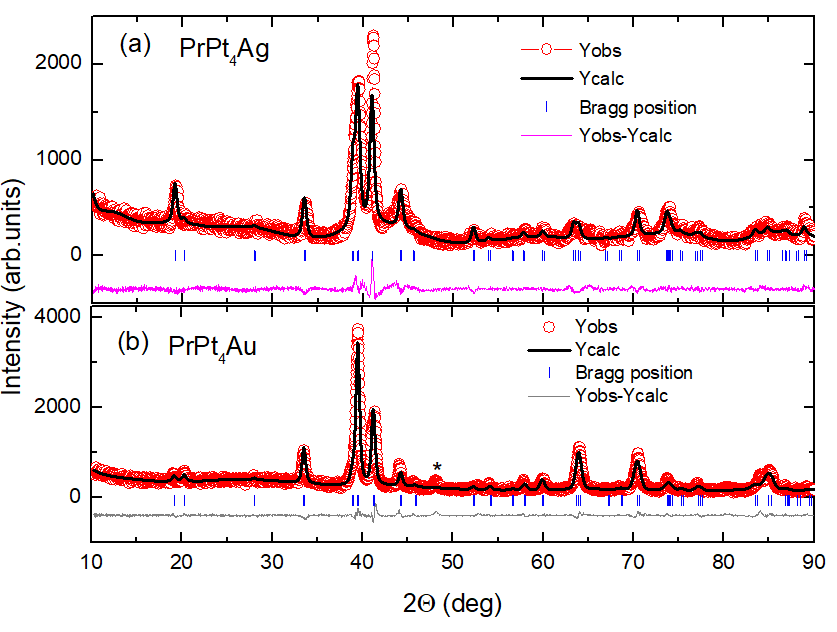}
	\caption{\label{fig_PrPt4X_XRD} Powder X-ray diffraction patterns of PrPt$_4$Ag and PrPt$_4$Au shown in (a) and (b) with Rietveld refinements (black lines) based on the $P$6/$mmm$  space group. The vertical bars are the Bragg peak positions, while the difference between the experimental and calculated intensities are shown as pink and gray lines, respectively. The peak in the PrPt$_4$Au spectrum marked with an asterisk does not belong to the major phase as discussed in the text.}
\end{figure}

Polycrystalline samples of PrPt$_4$Ag and PrPt$_4$Au were prepared by arc melting stoichiometric amounts of high-purity elements (wt.\% $\ge$ 99.9) on a water-cooled Cu plate under a purified static argon atmosphere in an Edmund Buehler arc furnace. The weight losses for both compounds were $\sim$ 0.09\%. The arc-melted pellets were then  wrapped in Ta foil, placed in an evacuated quartz tube and annealed at 800$^\circ$C for 14 days so as to improve the quality of the samples. Room temperature powder X-ray diffraction (XRD) patterns were recorded on pulverized samples using a Rigaku diffractometer employing Cu-K$\alpha$ radiation. The XRD patterns with the Rietveld refinements \cite{thompson1987rietveld} employing the FullProf suite of programs \cite{rodriguez1993recent} are shown in Fig.~\ref{fig_PrPt4X_XRD}. The lattice parameters obtained from the Rietveld refinements are presented in Table~\ref{PrPt4X_lattice_parameters}. For both compounds, the XRD results confirm phase formation of the desired compounds. For the case of PrPt$_4$Au, a spurious peak, barely resolved above the background noise was detected at 2$\Theta$ = 48$^{\circ}$. It was verified that this peak does not have its origin in any of the elements Pr, Pt or Au thus the origin of this peak remains unknown. \\
Magnetic properties were measured using the Magnetic Property Measurement System (Quantum Design Inc., San Diego) between 1.8~K and 300~K with an external magnetic field up to 7~T. The four-probe DC electrical resistivity, specific heat and thermal transport between 1.9~K and 300~K were measured using the Physical Property Measurement System also from Quantum Design.
\begin{table}[t!]
	\centering

	\caption{Lattice parameters of PrPt$_4$Ag and PrPt$_4$Au obtained from the Rietveld refinements of their XRD patterns. }
	\label{PrPt4X_lattice_parameters}
	\begin{tabular}{|c|c|c|c|c|c|c|}
		\hline
		Compound & $a$ (\AA) & $c$ (\AA) & $V$ (\AA$^3$) & $R_p$ (\%) & $R_{wp}$(\%) & $\chi^2$ \\ \hline
		
		PrPt$_4$Ag & 5.344(2) & 4.399(2) & 108.8(9) & 9.597 & 12.25 & 4.596\\
		PrPt$_4$Au & 5.343(3)&  4.378(5) &108.3(8) & 8.957&11.89& 4.875\\ \hline

	\end{tabular}
	
\end{table}
%
%
%
%

\section{Magnetic properties}
\label{magnetic}
\begin{figure}[!t]
	\centering
	\includegraphics[scale=0.7]{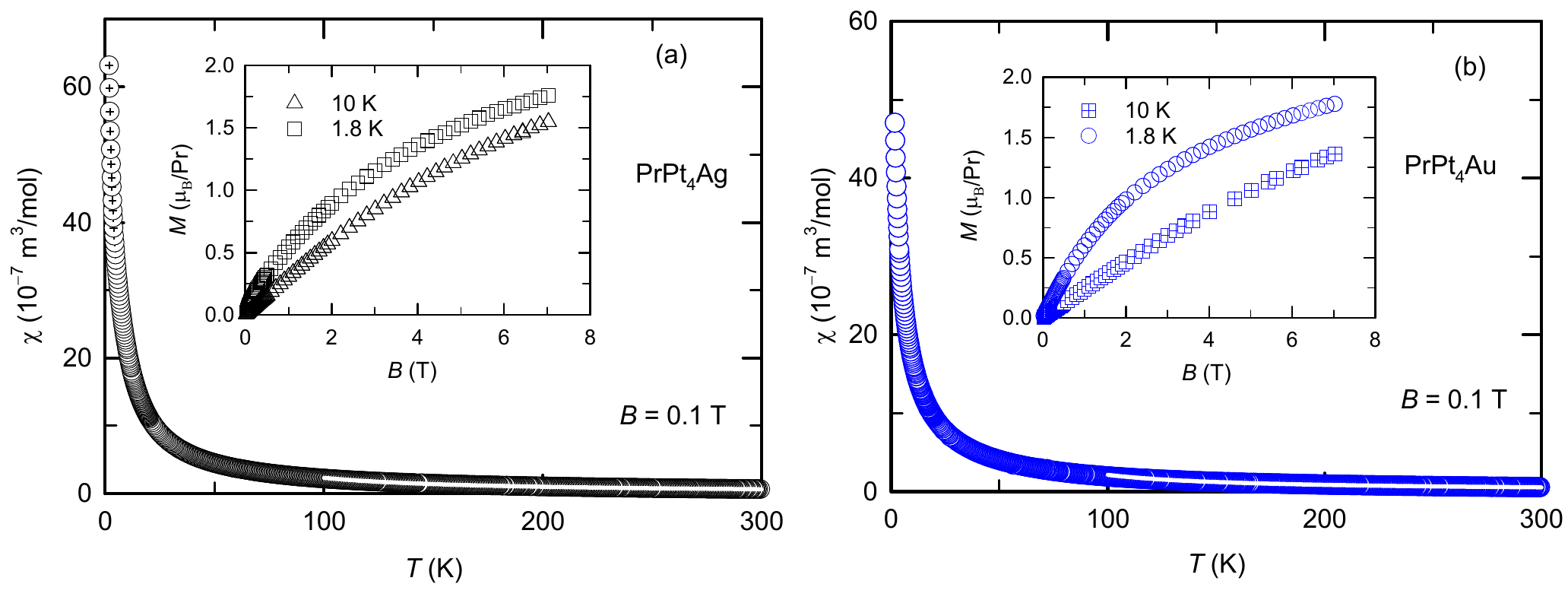}
	\caption{\label{fig_PrPt4X_chi} (a) Temperature dependence of magnetic susceptibility $\chi(T)$ of PrPt$_4$Ag in an external field of 0.1~T with a Curie-Weiss fit shown as a white line. Inset: Isothermal magnetization of  PrPt$_4$Ag at 1.8~K and 10~K. (b) Plot of $\chi$ against $T$ for PrPt$_4$Au with a Curie-Weiss fit represented by the white line. Inset: Isothermal magnetization of PrPt$_4$Au at 1.8~K and 10~K.}
\end{figure}

The temperature dependences of magnetic susceptibility $\chi(T)$ of both compounds in an external field of 0.1~T are shown in Fig.~\ref{fig_PrPt4X_chi} (a) and (b). From the plots, $\chi(T)$ of both compounds are qualitatively similar and show no visible anomaly down to 1.9~K.  For temperatures above 100~K, $\chi(T)$ for both compounds follow a Curie-Weiss behavior based on the expression: $\chi(T) = N_A \mu_\mathrm{eff}^2/(3k_B(T-\theta_p))$, where $\mu_\mathrm{eff}$ is the effective magnetic moment, $\theta_p$ is the Weiss temperature, N$_A$ is the Avogadro number and k$_B$ is the Boltzmann constant. From the least-square fits shown as white lines in Fig.~\ref{fig_PrPt4X_chi} (a) and (b), values of $\mu_\mathrm{eff} = 3.30\mu_B$/Pr, $\theta_P = 95.10~$K and  $\mu_\mathrm{eff} = 3.15~\mu_B$/Pr, $\theta_P = 99.16~$K for PrPt$_4$Ag and PrPt$_4$Au, respectively are obtained. These values of $\mu_\mathrm{eff}$ are slightly reduced in comparison to the value of 3.58~$\mu_B$/Pr calculated for a free Pr$^{3+}$ ion. Such reduction in $\mu_\mathrm{eff}$ can be attributed to the crystalline electric field (CEF) effect on the Pr$^{3+}$ moment. The isothermal magnetization for both compounds measured at temperatures of 1.8~K and 10~K are shown in the insets to Fig.~\ref{fig_PrPt4X_chi} (a) and (b).  The magnetization trend shows a little curvature at 10~K which becomes more pronounced at 1.8~K for both compounds. PrPt$_4$Ag and PrPt$_4$Au attain magnetization of $\approx$ 1.8~$\mu_B$/Pr at 7~T, which is about 50\% reduced compared to the value of 3.2~$\mu_B$/Pr expected for a free Pr$^{3+}$ ion. Assuming that at a temperature of 1.8~K no higher-lying levels of the $J$ = 4 multiplet of Pr are occupied, we attribute this deficiency in the extracted magnetization to the effects of magneto-crystalline anisotropy in both compounds.
\section{Specific heat}
\label{specific_heat}
\begin{figure}[!t]
	\centering
	\includegraphics[scale=0.7]{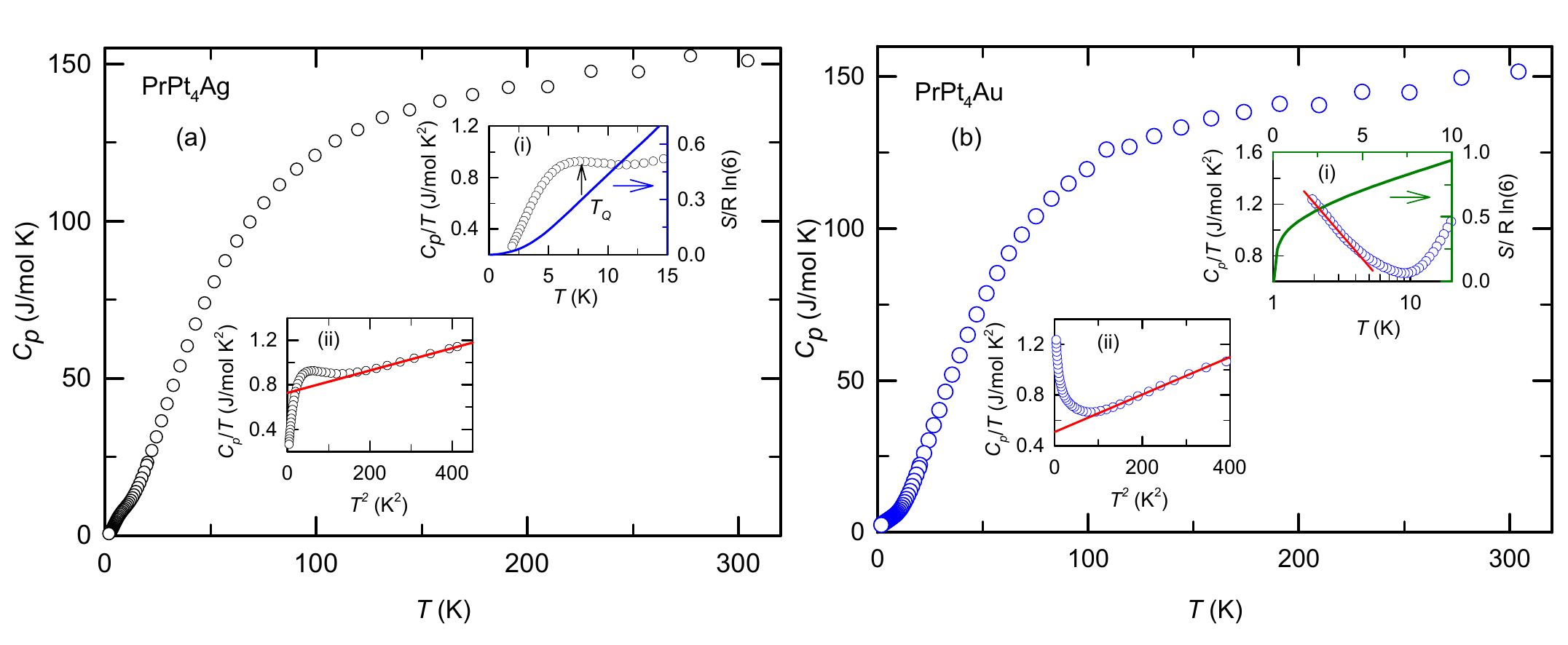}
	\caption{\label{fig_PrPr4Ag_cp} (a) Temperature dependence of specific heat $C_p(T)$ of PrPt$_4$Ag. Inset (i): Low-$T$ plot of $C_p/T$ against $T$ with the arrow indicating the phase transition at $T_Q$. Also shown is the calculated entropy (blue line) with axis on the right. Inset (ii): Plot of $C_p/T$ against $T^2$ for PrPt$_4$Ag together with a linear fit (red line) to extract the Sommerfeld coefficient. (b) Plot of $C_p$ against $T$ for PrPt$_4$Au. Inset (i): Plot of $C_p/T$ against $T$ on a semi-log axis. The red line is a guide to the eye indicating a logarithmic divergence of $C_p(T)/T$. The calculated entropy (green line) is also shown with axis on the right. Inset (ii): Plot of $C_p/T$ against $T^2$ for PrPt$_4$Au together with a linear fit (red line) to extract the Sommerfeld coefficient. }
\end{figure}

The temperature dependence of the specific heat $C_p(T)$ between 1.9~K and 300~K for PrPt$_4$Ag and PrPt$_4$Au are shown in Fig.~\ref{fig_PrPr4Ag_cp} (a) and (b). Values of $\approx$ 150~J/(mol K) are observed at room temperature for both compounds, which correspond to the Dulong-Petit value. In inset (i) of Fig.~\ref{fig_PrPr4Ag_cp} (a), the low-$T$ plot of $C_p/T$ against $T$ for PrPt$_4$Ag is shown. At 7.6~K an anomaly with a broad feature is observed. The electrical resistivity results to be presented in Section~\ref{resistivity} also reveal an anomaly around the same temperature. No such an anomaly is however observed in $\chi(T)$ presented in the previous section, thus suggesting a possible multipolar ordering of the Pr$^{3+}$ moment in PrPt$_4$Ag. This is supported by the Van-Vleck paramagnetism observed in $\chi(T)$ indicating that spins or dipolar order do not play a detectable role in the anomaly. In the absence of a long-range magnetic order in Pr systems, the 4$f$-electrons orbitals are the only effective degrees of freedom and multipolar ordering of the Pr$^{3+}$ moment can be realized \cite{suzuki2005quadrupolar,tanida2006possible}. Also shown is the plot of the calculated entropy. At $T_Q$ only an amount of 0.3R$\ln(6)$ is released which is only about one third of the total entropy expected. The plot of $C_p/T$ against $T$ on a semi-log scale for PrPt$_4$Au shown in inset (i) of Fig.~\ref{fig_PrPr4Ag_cp} (b) shows an upturn below 10~K reaching a value of 1.23~J/(mol K$^2$) at 1.9~K. We have also observed a linear-in-$T$ behavior of $\rho(T)$  which is discussed in Section~\ref{resistivity}. These observations in PrPt$_4$Au are the hallmark of a non-Fermi liquid (nFL) behavior. Among many other reasons, proximity to a quantum critical point or possible spin/elemental disorder such as in doped systems could be responsible for such behavior \cite{schofield1999non}. The calculated entropy is also presented in the same plot. At 8~K, the recovered entropy is about 0.9R$\ln(6)$ which is close to the full entropy expected.\\
\indent
We further analyzed the low-$T$ behavior by plotting of $C_p/T$ against $T^2$ for both compounds as shown in inset (ii) of Fig.~\ref{fig_PrPr4Ag_cp} (a) and inset (ii) of Fig.~\ref{fig_PrPr4Ag_cp} (b). Least-square fits to both sets of data (represented as red lines) based on the expressions: $C_p/T = \gamma + \beta T^2$ and $\beta = 12\pi^4nR/(5\theta^3_D$), where $n$ and R are the number of atoms per formula unit and universal gas constant, respectively, $\gamma$ is the Sommerfeld coefficient and $\theta_D$ is the Debye temperature are shown. From the fits, $\gamma$ = 728.50~J/(mol K$^2$) and 509.10~J/(mol K$^2$) are obtained for PrPt$_4$Ag and PrPt$_4$Au, respectively. The observed $\gamma$ values for these two compounds are more than two orders of magnitude greater than what is expected for an ordinary metal and qualitatively similar to those of the heavy fermion (HF) systems, where significant enhancement of the quasiparticles mass have been observed \cite{ogunbunmi2018electronic}.

\section{Electrical transport}
\label{resistivity}
\begin{figure}[!t]
	\centering
	\includegraphics[scale=0.7]{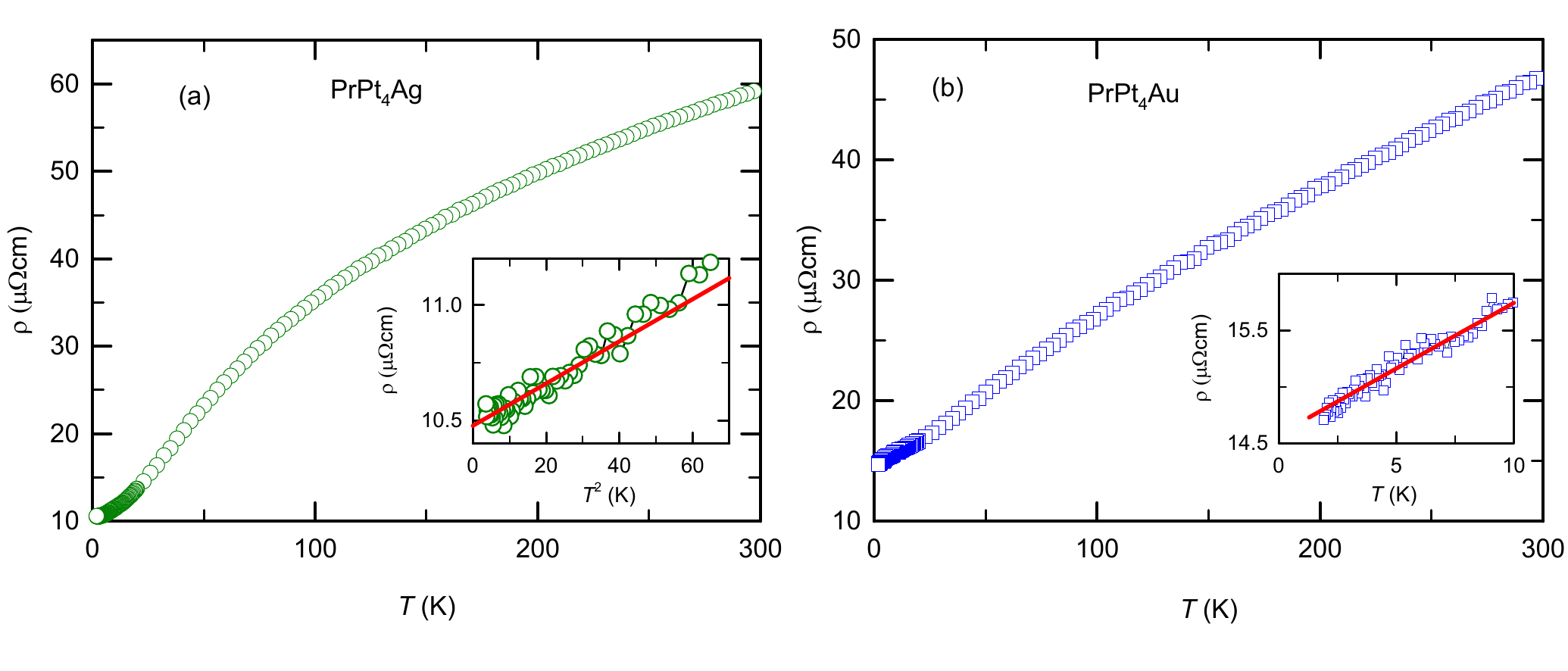}
	\caption{\label{fig_PrPt4Ag_rho} (a) Temperature dependence of electrical resistivity, $\rho(T)$, of PrPt$_4$Ag. Inset: Low-$T$ plot of $\rho$ against $T^2$ together with a fit (red line) described in the text. (b) Plot of $\rho(T)$ against $T$ of PrPt$_4$Au. Inset: Low-$T$ plot of $\rho$ and the red line is a fit described in the text.}
\end{figure}

The temperature dependence of the electrical resistivity $\rho(T)$ for PrPt$_4$Ag and PrPt$_4$Au investigated between 1.9~K and 300~K are presented in Fig.~\ref{fig_PrPt4Ag_rho} (a) and (b), respectively. The observed values of residual resistivity are $\approx$6 and $\approx$3 for PrPt$_4$Ag and PrPt$_4$Au, respectively. Both compounds show typical metallic behavior down to low temperatures. A broad curvature is observed at intermediate temperatures in PrPt$_4$Ag, which is common to rare-earth intermetallic systems; possibly associated to thermal de-population of the crystal field levels as temperature is lowered. A plot of $\rho(T)$ against $T^2$ presented in the inset of Fig.~\ref{fig_PrPt4Ag_rho} (a) indicates a power law behavior based on the expression: $\rho(T)$ = $\rho_0$ + $AT^2$. A fit of the expression to the data is shown as the red line. The residual resistivity, $\rho_0 = 10.46~\mu \Omega$cm and the coefficient of the quadratic term, $A = 0.010~\mu \Omega$cm/K$^2$ are obtained. The Kadowaki-Woods ratio (KWR) \cite{kadowaki1986universal}, $A/\gamma^2 = 1.88 \times 10^{-8}~\mu\Omega$cm(mol K/mJ)$^2$ is obtained by using $\gamma$ = 728.5~mJ/(mol K$^2$) obtained from $C_p(T)$ analysis. The relationship between $\chi(T)$ and $\gamma$ is also evaluated using the Wilson ratio \cite{wilson1975kg} given as: $R_w = \pi^2k_B^2\chi(T\longrightarrow 0)/(\mu^2_\mathrm{eff} \gamma)$. Using the observed low-$T$ values of $\chi(T\longrightarrow 0)$ = 0.5025~emu/mol, $\gamma$ = 728.5~mJ/(mol K$^2$) and $\mu_\mathrm{eff}$ = 3.30~$\mu_B$/Pr yields $R_w = 1.38$ which is comparable to a value of unity expected for HF systems. Also, a little anomaly is noticeable in the low-$T$ plot at $\sim$ 7.6~K  which coincides with the anomaly observed in $C_p(T)$ around the same temperature.\\
\indent
PrPt$_4$Au on the other hand is quasi-linear in nature down to low temperatures. In the inset of Fig.~\ref{fig_PrPt4Ag_rho} (b), the low-$T$ behavior of PrPt$_4$Ag is shown together with a fit (shown as red line) using the expression; $\rho(T)$ = $\rho_0$ + $AT^n$. From the fit, the residual resistivity, $\rho_0$ = 14.58~$\mu \Omega$cm, $n$ = 1 and $A$ = 0.117~$\mu \Omega$cm/K are obtained. Such a temperature dependence of $\rho(T)$ deviates from the Fermi-liquid behavior expected at low temperatures for normal metals. Results from $C_p(T)$ of PrPt$_4$Au discussed in Section~\ref{specific_heat} have also shown a temperature dependence that deviates from those of ordinary metals. Using the values of $A = 0.117~\mu \Omega$cm/K$^2$ and 509.10~mJ/(mol K$^2$), yields KWR = 4.49 $\times$ 10$^{-7}$~$\mu\Omega$cm(mol K/mJ)$^2$. Also, using the values of  $\chi(T\longrightarrow 0)$ = 0.3743~emu/mol, $\gamma$ = 509.10~mJ/(mol K$^2$) and $\mu_\mathrm{eff} = 3.15~\mu_B$/Pr yields $R_w$ = 1.62. The parameters observed in PrPt$_4$Ag and PrPt$_4$Ag reveal both compounds as new HF systems.
\begin{table}[t!]
	\centering
	\setlength{\tabcolsep}{0.5pt}
	
	\caption{A collection of estimated characteristic HF parameters of PrPt$_4$Ag and PrPt$_4$Ag.}
	\label{HF_properties}
	\begin{tabular}{|c|c|c|c|c|}
		\hline
		Compound & $\gamma$ (mJ/(mol$_\mathrm{Pr}$ K$^2$)) & $A$ ($\mu \Omega$cm/K$^2$) & $A/\gamma^2$ ( $\times$ 10$^{-5}$$\mu\Omega$cm(mol$_\mathrm{Pr}$ K/mJ)$^2$) & $R_w$  \\ \hline
		
		PrPt$_4$Ag  & 728.50 & 0.010 & 0.00188 & 1.38  \\ \hline
		PrPt$_4$Au  & 509.10&  0.117 &0.0449 & 1.62 \\ \hline

	\end{tabular}
	
\end{table}

\section{Discussion and conclusion}
The existence of the hexagonal PrPt$_4$Ag and PrPt$_4$Au compounds are reported together with their physical and magnetic properties. PrPt$_4$Ag shows a putative multipolar ordering at $T_Q = 7.6~$K. From the crystal field levels permitted for the $J$ = 4 multiplet in a $D_{6h}$ local symmetry, one would expect to have 3 $\Gamma_3$ non-Kramers doublets and another 3 $\Gamma_1$ singlets. However, the observation of such an anomaly at $T_Q$ makes $\Gamma_3$ the likely ground state in PrPt$_4$Ag. In contrast, the low-$T$ region of PrPt$_4$Au shows a logarithmic divergence of $C_p(T)/T$, and a linear-in-$T$ behavior of $\rho(T )$ below 10~K which is typical of a non-Fermi liquid behavior. These behaviors are likely associated with spin scatterings observed for a system in the proximity of a quantum critical point. Further analyses of $C_p(T)$ of both compounds reveal $\gamma$ = 728.50~J/(mol K$^2$) and 509.10~J/(mol K$^2$) for PrPt$_4$Ag and PrPt$_4$Au, respectively. In Table~\ref{HF_properties}, a collection of characteristic HF parameters are presented for easy comparison with the observations in this work. The observation of enhanced $\gamma$ value in systems with a $\Gamma_3$ non-Kramers ground state are often associated to the quadrupolar Kondo effect \cite{suzuki2005quadrupolar,tanida2006possible}. Future studies will focus on experiments below 2~K in order to further unravel the physics involved in the ground state of these two non-Kramers-ion systems.
\section*{Acknowledgment}
MOO acknowledges the UJ-URC bursary for doctoral studies in the Faculty of Science. AMS thanks the SA-NRF (93549) and UJ-URC for financial support.

\bibliographystyle{iopart-num}
\bibliography{PrPt4X}

\end{document}